\documentstyle[aps,epsfig,multicol,amstex,subfigure]{revtex}
\draft
\begin{document}
\title{Non-Equilibrium Electron Transport in Two-Dimensional
Nano-Structures Modeled by Green's Functions and the Finite-Element
Method} \author{P. Havu$^1$, V. Havu$^2$, M. J. Puska$^1$, and
R. M. Nieminen$^1$} \address{ 1) Laboratory of Physics, Helsinki
University of Technology, P.O. Box 1100, FIN-02015 HUT, Finland \\ 2)
Institute of Mathematics, Helsinki University of Technology, P.O. Box
1100, FIN-02015 HUT, Finland}

\maketitle
\begin{abstract}
We use the effective-mass approximation and the density-functional theory
with the local-density approximation for modeling two-dimensional
nano-structures connected phase-coherently to two infinite
leads. Using the non-equilibrium Green's function method the
electron density and the current are calculated under a bias voltage.
The problem of solving for the Green's functions numerically is formulated
using the finite-element method (FEM). The Green's functions have
non-reflecting open boundary conditions to take care of the infinite
size of the system.  We show how these boundary conditions are
formulated in the FEM. The scheme is tested by calculating
transmission probabilities for simple model potentials. The potential
of the scheme is demonstrated by determining non-linear current-voltage
behaviors of resonant tunneling structures.
\end{abstract}


\begin{multicols}{2}
%-------------------------------------------------------------------
\section{Introduction}

Two-dimensional (2D) nanodevices are structures in which electrons
move in a restricted nanometer-size area. The phase-coherence length
of electrons is of the order of the dimensions of the device. Electron
transport through nanodevices cannot be modeled using the traditional
description based on diffusion or Boltzmann equations. One has to use
a method which takes the quantum-mechanical character of the carriers,
e.g. quantum interference, explicitly into account \cite{datta}.

Nanodevices are fabricated using semiconductor-heterostructure
techniques. A layer of semiconductor (e.g. AlGaAs) is grown on top of
another semiconductor (GaAs) with molecular-beam epitaxy. The two
semiconductors have different band gaps so that electrons accumulate
in the potential well at the semiconductor interface and form a 2D
electron gas. Above of the semiconductor layer metallic gates are
fabricated. Applying voltage on them the electron motion can also be
restricted in the horizontal direction and nanodevices, such as
quantum point contacts and quantum dots, are created.

The quantum-mechanical modeling of 2D nanostuctures is usually based
on the effective-mass approximation. For the ground-state carrier
distribution one can employ, for example, Monte Carlo-methods
\cite{montecarlo} or density-functional theory (DFT)
\cite{manninen}. The description of isolated structures is rather
straightforward because the system is finite and all the electron
states can be calculated. Often the nanodevice is connected to a
measuring system by leads and the current through the system is
measured. If the connection is weak the nanostucture can still be
approximated as an isolated system, but in the case of strong coupling
the combined nanostructure-leads system has to be described.  In this
case the leads can have a considerable effect on the electronic
structure of the nanodevice.  The electronic structure of this kind of
open system can be obtained using DFT by calculating the wave
functions in the scattering formalism using the Lippmann-Schwinger
equation \cite{lang}. The method also relates to the conductance of the
system in the limit of zero bias. Another possibility is to use DFT in
combined with the non-equilibrium Green's function (NEGF) method
\cite{green1}. In this scheme the wave functions are not calculated
explicitly in the device region. The NEGF-approach also enables the
addition of a bias voltage between the leads and the calculation of
the current through the system also in the non-equilibrium state.

The electronic-structure calculations using the Green's functions
demand extensive computer resources. Therefore the numerical method
for the Green's function implementation has to be chosen carefully.
There is a wide range of different numerical methods available today
for electronic structure calculations, e.g. the finite-difference
method \cite{beck}, the linear combinations of atomic orbitals (LCAO)
method, the wavelet method \cite{arias}, and the plane-wave method
\cite{uasp} among the most popular ones.  Previously, the Green's
function method coupled to DFT has been used in nanostucture
calculations employing atomic orbitals \cite{siesta1,siesta2},
localized optimized orbitals in real space \cite{bernholc}, Gaussian
orbitals \cite{damle} or wavelets \cite{wavelet} as basis functions.

In the present work we have adopted the finite-element method (FEM) to
study 2D nanostructures within the effective-mass theory and using the
DFT-NEGF scheme.  Previously, electronic structure calculations the
FEM has been used in, for example, in Refs.
\cite{fem1,fem2,fem3,fem4,fem5}.  The main advantages gained by the
FEM in the present context are the possibility to control the accuracy
of the approximation via mesh refinements, the ability to simulate
easily different geometrical configurations of the system and the ease
in the treatment of the boundary conditions. Moreover, the evaluation of
the basis functions is fast and the ensuing sparse linear systems
allow the use of fast sparse solvers. In practice, we have chosen to
use piecewise polynomials as basis functions.  The polynomials are
very fast and stable to evaluate in any computational environment. The
approximation properties of the polynomials are well-known and several
error bounds are available \cite{femKirja}.  In the FEM the open
boundary conditions are easier to implement than in the
finite-difference method \cite{datta} and in the basis set methods
\cite{siesta1,siesta2,wavelet} in which they are derived by first
writing down the infinite discretization matrix and then cutting out the
central area from it. In the FEM these boundary conditions are
written in a simpler and more intuitive way as will be shown in this
work.

We use use effective atomic units which are derived by putting the
fundamental constants $e = \hbar = m_e = 1$, and the material
constants, the effective electron mass and the dielectric constant
$m^* = \epsilon = 1$ respectively.  The effective atomic units are
transformed to the usual atomic units using the relations
\begin{center}
\begin{tabular}{l l l l}
Length: & $1 \, a_0^*$ & $= 1 \frac{\epsilon}{m^*} a_0$ & $\approx
\frac{\epsilon}{m^*} \, 0.529177 \times 10^{-10} $ m \\
Energy: & $1 \, {\rm Ha^*}$ &$= 1 \frac{m^*}{\epsilon^2} {\rm Ha}$ &
$\approx \frac{m^*}{\epsilon^2} \, 27.2116$ eV \\ 
Current: & $1 \, {\rm a.u.}^*$&$ = 1 
\frac{m^*}{\epsilon^2} {\rm a.u.} $ &
$\approx \frac{m^*}{\epsilon^2} \, 6.6231$ mA.  
\end{tabular}
\end{center}

The organization of the present paper is as follows. In
Sec.~\ref{model}. we present our 2D nanostucture model and explain how
the Green's functions are used in the electronic structure and current
calculations. In Sec.~\ref{fem} we formulate the solution of the
Green's functions within the FEM. Finally, in Sec.~\ref{test} we deal
with our test cases, which include confining well and bottle-neck
model potentials and double-wall barrier systems. Sec. V contains the
conclusions.

%----------------------------------------------------------------- 
\section{Model and Green's function formulation}\label{model}

\subsection{The model for two-dimensional nanostructures}

In real nanodevices electrons of the 2D electron gas are in a
potential well at the interface between two semiconductors.  The
electron density in the well is neutralized by a positively-ionized
donor layer separated from the potential well. The lateral confinement
of electrons is obtained by gate voltages.  Electrons are in practice
in the ground-state with respect to the motion perpendicular to the
interface. Therefore our model is strictly two-dimensional.
 
\begin{figure}[htb]
\begin{center}
\epsfig{file=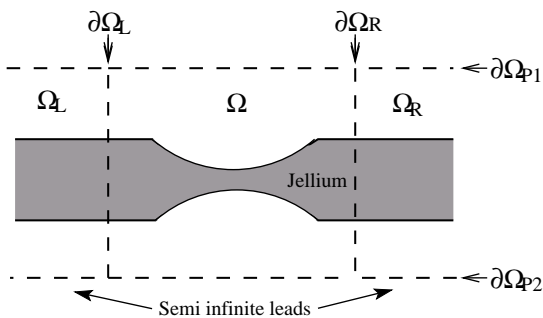,width=0.4\textwidth}
\end{center}
\caption{\label{dotti3} Model nanostructure between
two infinite leads.}
\end{figure}

A schematic sketch of the model is in Fig.~\ref{dotti3}. It shows the
region of interest between two semi-infinite leads.  The potential
profile is a combination of interactions between electrons and the
positive constant background charge (jellium), and the external
potential caused by the gate voltages. Thus, the layer of ionized
donors and the 2D electron layer coincide in our model.  In many
models the potential profile is approximated using a harmonic
potential profile \cite{manninen,hirose}. In our model this
approximation cannot be used, because we solve for the electrostatic
potential of an infinite system requiring that the system is charge
neutral.  In order to keep the model simple the confinement of the
electrons is established by shaping the background charge and,
optionally, by external potentials in certain regions of the system.

We divide the infinite system to three separate areas as shown in
Fig.~\ref{dotti3}, the central area $\Omega$, the left region
$\Omega_L$ and the right region $\Omega_R$. We denote the boundary
between the regions $\Omega$ and $\Omega_R$ as $\partial \Omega_R$ and
between the regions $\Omega$ and $\Omega_R$ as $\partial
\Omega_R$. The Green's functions are calculated in the region
$\Omega$.  $\partial \Omega_L$ and $\partial \Omega_R$ are
non-reflecting open boundaries. On the other two boundaries, $\partial
\Omega_{P1/P2}$ which are far enough from the important device region
the potential is assumed to be infinite, so that the Green's functions
vanish there.

We solve for the self-consistent electron structure of the system
iteratively. The electron density is calculated from the Green's
functions. The effective potential is calculated from the electron
density as usual in the DFT within the local-density approximation
(LDA). After mixing the new effective potential with potential from
the previous iteration the electron density is recalculated. The loop
is repeated until convergence is achieved.

The effective potential has four terms
\begin{equation}
V_{eff} = V_c + V_{xc} + V_{bias} + V_{gate},
\end{equation}
where $V_c$ and $V_{xc}$ are the Coulomb and the exchange-correlation
potentials arising from the charge distributions, respectively. The
calculation of $V_c$ is discussed below in more detail. For $V_{xc}$,
we use the recent 2D-LDA functional by Attacalite et al. \cite{xc1,xc2}.

\begin{figure}[htb]
\begin{center}
\epsfig{file=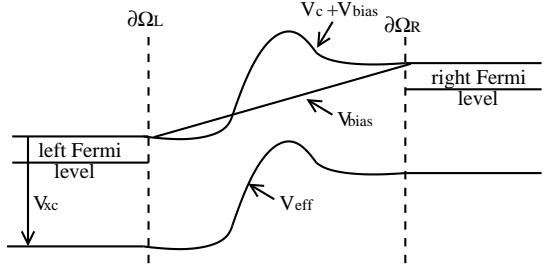,width=0.4\textwidth}
\end{center}
\caption{\label{sivukuva} Effective potentials and Fermi levels under
  the bias voltage.}
\end{figure}

$V_{bias}$ takes care of the boundary conditions under the bias
voltage \cite{green1}. The total electrostatic potential has different
levels in the right and left leads. This introduces $V_{bias}$ as a
linear ramp potential over $\Omega$. In the regions $\Omega_L$ and
$\Omega_R$, $V_{eff}$ is calculated as a potential of the infinite
(jellium) wire. Then $V_{eff}$ is also continuous if $\Omega$ is large
enough.  The ensuing energy scheme is shown in
Fig.~\ref{sivukuva}. Also the Fermi levels in the right and left leads
differ by the applied bias voltage $\Delta V_{bias}$.  $V_{gate}$ is
an external gate potential. Using gate voltages it is possible to
increase or decrease the potential in certain regions, for example to
increase the potential walls and to decrease the potential wells of a
bare jellium system.

Below we use a notation in which a point inside the
two-dimensional region $\Omega$ is denoted by $r$ and a point outside
the region $\Omega$ in region $\Omega_R$ or $\Omega_L$ by $r_e$. A
point on the boundary $\partial \Omega_L$ is $r_L$ and a point on
$\partial \Omega_R$ is $r_R$.

%------------------------------------------------------------------
\subsection{Green's functions in electronic-structure calculations}

\begin{figure}
  \centering
  \mbox{\subfigure{\epsfig{figure=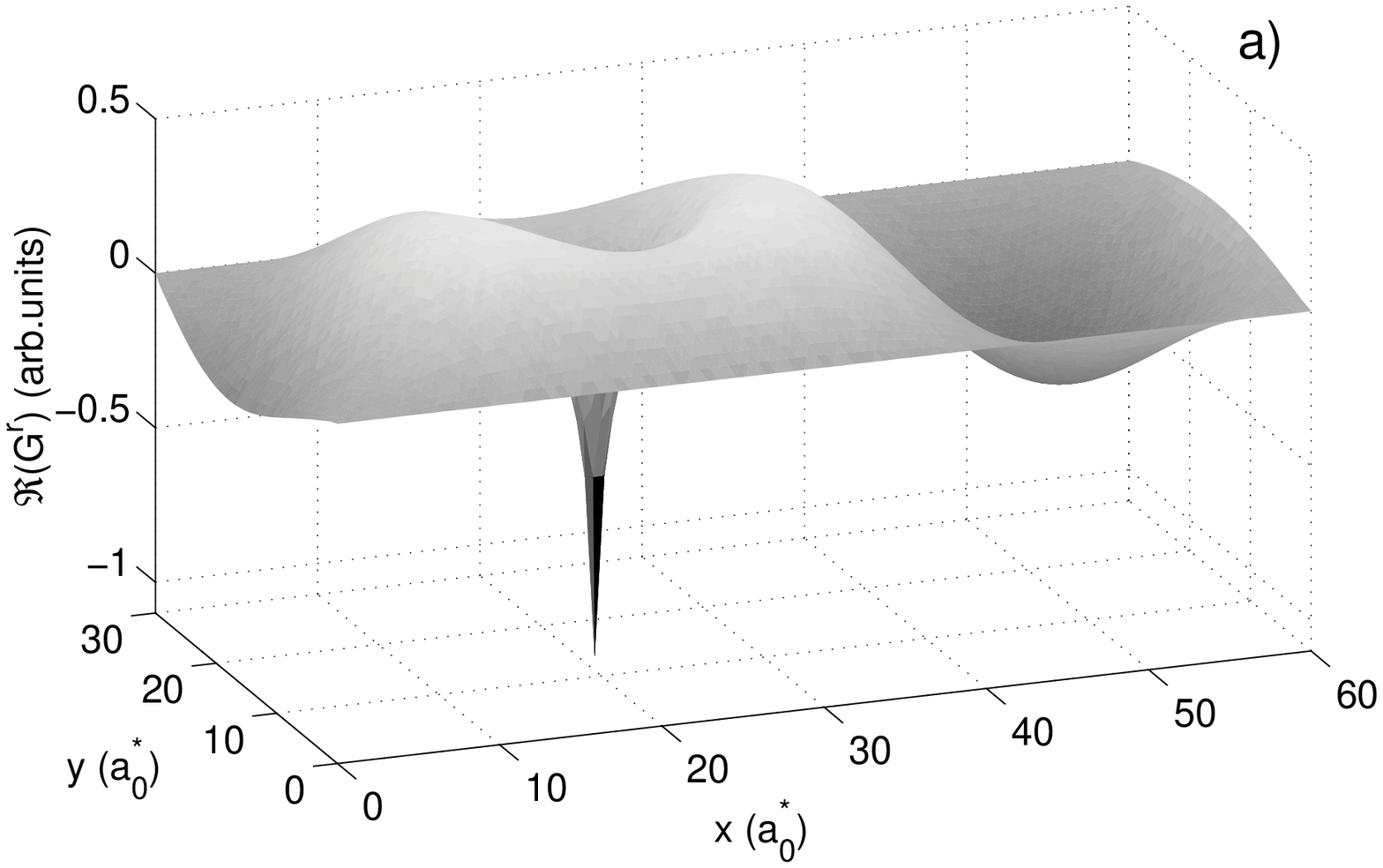,
        width=0.4\textwidth}} } \\
  \mbox{
    \subfigure{\epsfig{figure=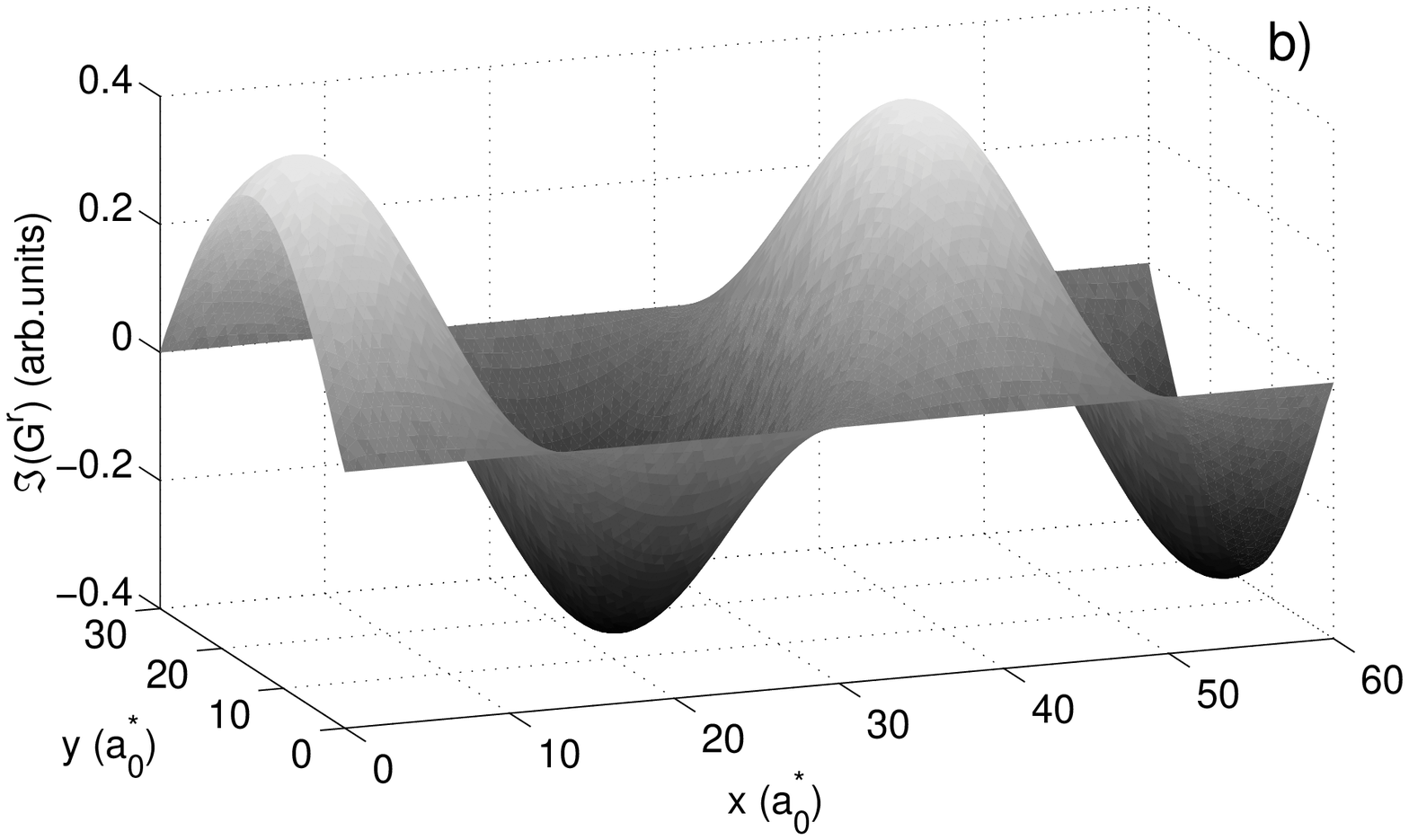,
        width=0.4\textwidth}} }
  \caption{Real (a) and imaginary (b) parts of the Green's function
  $G^r(r,r')$ for a uniform jellium wire. $r \, = \, (x,y)$ and $r'$ =
  (21.6, 15.4) (the position of the pole). }
  \label{greenRKuva}
\end{figure}

We use Green's functions in calculating the electronic structure and
the current under an external bias voltage. The theory is explained in
more detail in Refs.~\cite{datta} and \cite{green1}. The electron
density is calculated from the Green's function $G^<$. In order to obtain
$G^<$ one has  to solve first for the retarded Green's function $G^r$
from
\begin{equation}\label{greenR}
\big{(}\omega-\hat{H}(r)\big{)} G^r(r,r';w) = \delta(r-r'),
\end{equation}
where $\omega$ is the electron energy and $\hat{H}$ is the DFT Hamiltonian
the system,
\begin{equation}
\hat{H}(r) = -\frac{1}{2}\nabla^2 + V_{eff}(r).
\end{equation}
In this case $r$ is a two-dimensional variable. Its components along and
perpendicular to the leads are $x$ and $y$, respectively. $G^r$ is
zero on the boundaries parallel to the leads (see
Fig.~\ref{dotti3}). If $\omega$ is smaller than the bottom of the
potential $V_{eff}$ in the lead Eq.~(\ref{greenR}) gives
exponentially decaying solutions there. Otherwise the solution
oscillates with a constant amplitude to the infinity. The form of
$G^r(r,r')$ in a uniform jellium wire is shown in
Fig~\ref{greenRKuva}. The real part has a pole at $r=r'$, while the
imaginary part behaves smoothly everywhere. This is why the imaginary
part is much easier to approximate numerically than the real
part.

In
equilibrium, when the Fermi functions in $\Omega_L$ and $\Omega_R$ are
identical, $f_L(w) \equiv f_R(w) $, we obtain
\begin{equation}\label{ele1}
G^<(r,r';w) = 2f_{L/R}(w) G^r(r,r';w).
\end{equation}
This equation is also valid under a bias voltage at energies $\omega$
for which $f_L(\omega) = f_R(\omega)$ (in practice, $f_{L/R}=1$ for
those energies). If Eq.(\ref{ele1}) is not applicate, $G^<$ has to be
calculated in a more complicated way. Eq.  (\ref{greenR}) can be
reformulated using the so-called retarded self-energies of the leads,
$\Sigma_R^r$ and $\Sigma_L^r$, as
\begin{equation}
\big{(}\omega-\hat{H_0}-\Sigma_L^r(\omega) - \Sigma_R^r(\omega)
\big{)} G^r(r,r';w)
= \delta(r-r').
\end{equation}
Above, $\hat{H_0}$ is the Hamilton operator for the isolated central
area $\Omega$. In practice, $\Sigma_{L/R}$ can be calculated from the
boundary conditions for the Green's functions at $\partial
\Omega_{L/R}$.  $\Sigma_{L/R}$ are functions with non-zero values only
at the boundaries $\partial \Omega_{L/R}$. Next we define the
functions $\Gamma_{L/R}$ as
\begin{equation}\label{gammat}
\begin{aligned}
%\begin{eqnarray}
i\Gamma_L &= \Sigma^r_L - \Sigma^a_L = 2i \Im(\Sigma^r_L),
\\ i\Gamma_R &= \Sigma^r_R - \Sigma^a_R = 2i \Im(\Sigma^r_R).
\end{aligned}
\end{equation}
%\end{eqnarray}
%
$\Sigma^a_{L/R}$ are the self-energies for the advanced Green's
function $G^a = (G^r)^*$. One can then write the electron density as the
sum of the electron flows from the leads to the region $\Omega$, using
\begin{equation}
\begin{aligned}
\label{ele2}
 G^<&(r,r';\omega) = \\
& -i f_R(\omega) \int_{\partial \Omega_R} \int_{\partial \Omega_R} G^r(r,r_R;\omega)
 \Gamma_R(r_R,r'_R;\omega) \\
  & \quad \quad \quad \quad \quad \quad \quad \times
 G^a(r'_R,r';\omega) \,
dr_R \, dr'_R  \\ 
 &-if_L(\omega)  \int_{\partial \Omega_L}\int_{\partial \Omega_L}
G^r(r,r_L;\omega) \Gamma_L(r_L,r'_L;\omega) \\
 & \quad \quad \quad \quad \quad \quad  \quad \times
G^a(r'_L,r';\omega) \,
 dr_L \, dr'_L,
\end{aligned}
\end{equation}
where $f_{R/L}$ are the Fermi functions in the right and left leads.
This equation has to be used in nonequilibrium situations when $f_R
\neq f_L$.

Eq.~(\ref{ele2}) corresponds to the electron density due to the states
extending to infinity in the leads.  Eq.~(\ref{ele1}) includes also the
electron density of possible bound states, which are localized near
$\Omega$ and decay exponentially in the leads.

\begin{figure}[htb]
\begin{center}
\epsfig{file=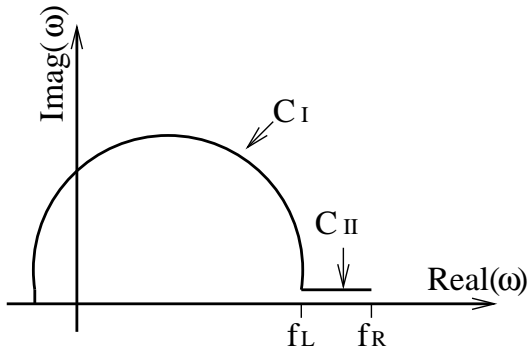,width=0.4\textwidth}
\end{center}
\caption{\label{energiaInt} Integration path used in
Eq.~(\ref{ele_integraali}). }
\end{figure}

In order to calculate total electron density we integrate over
the electron energy~$\omega$
\begin{equation}\label{ele_integraali}
\rho(r) = \frac{-1}{2\pi} \int_{-\infty}^{\infty} \Im(G^<(r,r;\omega)) d\omega.
\end{equation}
We use both equations (\ref{ele1}) and (\ref{ele2}) in this
integration.  Eq.~(\ref{ele1}) is analytic in the upper half of the
imaginary $\omega$-plane whereas Eq.~(\ref{ele2}) has poles below and
above the real $\omega$ axis. Thus, using Eq.~(\ref{ele1}) it is
possible the transfer the integral path from the real axis to the
complex plane. Our integration path is shown in Fig
\ref{energiaInt}. The first part is a semi-circle $C_I$ in the complex
$\omega$-plane using Eq. (\ref{ele1}) and it takes care of the
possible bound states below the energy bands of the leads. The rest of
the integration, $C_{II}$, is close to the real axis and there
Eq. (\ref{ele2}) is used. On the semi-circle only few integration
points are needed because the rapid variations of $G^<$ are smeared
out when the integration leaves the real axis. This is specially
useful for the bound states, which give rise to sharp peaks near the
real axis.

Computationally, it is faster to solve for $G^<$ from Eq.~(\ref{ele2})
than from Eq.~(\ref{ele1}).  Eq.~(\ref{ele1}) results in the inversion
of the entire matrix, because one needs $G^r(r,r')$ in all the
discretion points of $\Omega$. Electron density in
Eq.~(\ref{ele_integraali}) is the calculated using the diagonal
entries of the imaginary part, $\Im(G^r(r,r))$.  Inversion of the
matrix using direct sparse routines from HSL \cite{hsl} occurs as
follows. First one performs the symbolic analysis and factorization to
produce an ordering that reduces the fill-in. After that a numerical
factorization with pivoting is performed producing the Cholesky factor
of the matrix. The a set of linear equations with different right-hand
sides are solved. The number of equation is equal to the dimension of
the matrix.  Eq.~(\ref{ele2}) needs only the Green's functions
$G^r(r,r')$ for $r'=r_{L/R}$ on the boundaries $\partial
\Omega_{L/R}$.  This means that after factorization one has to solve
for a set of only as many linear equations as there are discretization
points on $\partial \Omega_{L/R}$.

For 2D systems the use of Eq.~(\ref{ele1}) is justified because the
analytic continuation of the integrand reduces the number of points
needed in the numerical integration of Eq.~(\ref{ele_integraali}) and
because the discretization error is smaller for
Eq.~(\ref{ele1}) than for Eq.~(\ref{ele2}). Namely, only the imaginary
part of $G^r$ is used in Eq.~(\ref{ele_integraali}) so that the pole
of $\Re(G^r)$ does not cause any major numerical problems if
Eq.~(\ref{ele1}) is used.

%-------------------------------------------------------------

\subsection{Electric Current}

The electric current is also calculated using the Green's functions.
The electron tunneling probability through the central region is obtained
from
\begin{equation}\label{tunneling_probability}
\begin{aligned}
T(\omega) =& \int_{\partial \Omega_L} \int_{\partial \Omega_L}
\int_{\partial \Omega_R}\int_{\partial \Omega_R} 
\Gamma_L(r_L,r'_L;\omega) 
 G^r(r'_L,r_R;\omega ) \\
\times& \Gamma_R(r_R,r'_R;\omega)
G^a(r'_R,r_L;\omega)
\, dr_L \, dr'_L \, dr_R \,dr'_R,
\end{aligned}
\end{equation}
and the total current is calculated integrating over the energy,
$\omega$ and taking care of the electron occupations in both leads. In
the effective atomic units the result is
\begin{equation}
I = \frac{1}{\pi} \int_{-\infty}^{\infty} T(\omega) \left(
f_{L}(\omega) - f_R(\omega) \right) d\omega.
\end{equation}

%---------------------------------------------------------------
\section{Finite-element method for solving Green's functions}\label{fem}

\subsection{\label{sec:variational}Variational formulation}

The most demanding computational task is to find the Green's function
at different energies as presented above. To this end, we first divide
the domain of the problem into two disjoint parts, the computational
domain $\Omega$ and the exterior domain $\Omega^e$. Only the
computational domain is discretized whereas the exterior is taken care
of by the corresponding Green's function (see below
Ch.~\ref{greenulko}). First, we cast Eq.~(\ref{greenR}) into a
variational, or weak, formulation for the domain $\Omega$. During the
derivation we frequently make use of the Green's formula
\begin{equation}
  \label{eq:greens_formula}
  \int_\Omega \nabla u \cdot \nabla v \, dr = \int_{\partial \Omega}
  \frac{\partial u}{\partial n} v \, ds - \int_\Omega u \nabla^2 v \, dr,
\end{equation}
for two arbitrary, sufficiently smoothly behaving functions $u$ and
$v$.  Above, $n$ denotes the outward normal of $\Omega$, and the line
integration is taken in the counter-clockwise direction around the
2D area $\Omega$.

To proceed, we multiply Eq.~(\ref{greenR}) by a sufficiently
smooth function $v$ and integrate the resulting identity over $\Omega$
giving
\begin{equation}
  \begin{aligned}
    \int_\Omega & v(r)\big{[} \omega - \hat{H}(r) \big{]} G^r (r , r'
    ; \omega)\,dr\\
    =&\int_\Omega v(r)\Big{\{} \frac{1}{2}\nabla^2 G^r (r,r'; \omega)\\
    &+ \big{[}\omega- V_{eff}(r) \big{]} G^r(r,r'; \omega) \Big{\}} \, dr \\
    =&  \int_\Omega v(r) \delta(r-r') \, dr \\
    =& \, v(r'),
  \end{aligned}
\end{equation}
The use of the Green's formula of Eq.~(\ref{eq:greens_formula}) gives
\begin{equation}
  \begin{aligned}
    \int_\Omega & v(r) \frac{1}{2}\nabla^2 G^r(r,r'; \omega) \, dr \\
    =&  -\int_\Omega  \nabla v(r)  \cdot \frac{1}{2} \nabla G^r(r,r'; 
    \omega) \, dr \\
    & + \int_{\partial \Omega_L} v(r_L)\frac{1}{2} \frac{\partial 
      G^r(r_L,r'; \omega)}{\partial n_L} \, dr_L \\
    & + \int_{\partial \Omega_R} v(r_R) \frac{1}{2} \frac{\partial G^r(r_R,r';
      \omega)}{\partial n_R}  \, dr_R.
  \end{aligned}
\end{equation}
Thus, the original problem of Eq.~(\ref{greenR}) is equivalent to the
formulation
\begin{equation}
  \label{eq:variational_formulation}
  \begin{aligned}
    \int_\Omega &\Big{\{}  - \nabla v(r) \cdot \frac{1}{2} \nabla
    G^r (r,r';\omega)  \\
   & + v(r) \big{[} \omega-V_{eff}(r) \big{]} G^r(r,r' ; \omega) \Big{\}} \,dr\\
   &  + \int_{\partial \Omega_L} \frac{1}{2} \frac{\partial
      G^r(r_L,r';
      \omega)}{\partial n_L} v(r_L) \, dr_L \\
    & + \int_{\partial \Omega_R} \frac{1}{2} \frac{\partial
      G^r(r_R,r'; \omega)}{\partial n_R} v(r_R) \, dr_R \\
    =& \, v(r')
  \end{aligned}
\end{equation}
for any sufficiently smooth function $v$.

In order to obtain a solvable system, the boundary conditions must be
supplied at the boundaries $\partial \Omega_L$ and $\partial
\Omega_R$. For conciseness we discuss only the case of $\partial
\Omega_L$, the other case $\partial \Omega_R$ being similar. Consider
the exterior problem
\begin{eqnarray}
  \big{(} \omega - \hat{H}(r_e) \big{)} g_e (r_e, r'_e ; \omega) & = & 
  \delta(r_e - r'_e), \quad r'_e \in \Omega_L \nonumber \\
  g_e (r_e, r'_e ; \omega) & = & 0, \quad r_e \in \partial \Omega_L,
\end{eqnarray}
for the Green's functions $g_e$ of the semi-infinite lead.  It follows
that any sufficiently smooth function $u$ can be written in the form
\begin{equation}
  \begin{aligned}
    u(r'_e) =&  \int_{\Omega_L} u(r_e) \delta(r_e - r'_e) \, dr_e  \\
    = &\int_{\Omega_L} u(r_e) \big{[} \omega - \hat{H}(r_e)\big{]} g_e (r_e, r'_e;
    \omega) \, d r_e \\ 
    = &\int_{\Omega_L} u(r_e) \Big{\{} \frac{1}{2}
    \nabla^2 g_e (r_e, r'_e; \omega) \\ 
    & + \big{[}\omega-V_{eff}(r)\big{]} g_e (r_e,
    r'_e; \omega) \Big{\}} \, d r_e
  \end{aligned}
\end{equation}
for $r'_e \in \Omega_L$. Using the Green's formula
(\ref{eq:greens_formula}) for the exterior domain $\Omega_L$ twice we
can write
\begin{equation}
  \begin{aligned}
    \int_{\Omega_L} & u(r_e) \frac{1}{2} \nabla^2 g_e(r_e, r'_e ;
    \omega) \, d r_e \\ 
    =& -\int_{\Omega_L} \frac{1}{2} \nabla u(r_e)
    \cdot \nabla g_e(r_e, r'_e ; \omega) \, d r_e \\ 
    &+ \int_{\partial
    \Omega_L} \frac{1}{2} u (r_L') \frac{\partial g_e(r_L', r'_e ;
    \omega)}{\partial n'_L} \, d r_L' \\ 
    =& \int_{\Omega_L}
    \frac{1}{2} g_e(r_e,r'_e; \omega) \nabla^2 u(r_e) \, d r_e \\ 
    &+ \int_{\partial \Omega_L} \frac{1}{2} u (r_L') \frac{\partial
    g_e(r_L', r'_e ; \omega)}{\partial n_L'} \, d r_L' \\ 
    & -\int_{\partial \Omega_L} \frac{1}{2} \frac{\partial
    u(r_L')}{\partial n_L'} g_e(r_L', r'_e ; \omega) \, d r_L',
  \end{aligned}
\end{equation}
so that
\begin{equation}
  \label{eq:represe}
  \begin{aligned}
  u(r'_e) =& \int_{\Omega_L} g_e(r_e,r'_e;\omega) \big{[} \omega -
 \hat{H}(r_e) \big{]} u(r_e) \, d r_e  \\  
 &+\int_{\partial \Omega_L} \frac{1}{2} u (r_L')
    \frac{\partial g_e(r_L', r'_e ; \omega)}{\partial n_L'} \, d r_L' \\
   & - \int_{\partial \Omega_L} \frac{1}{2} \frac{\partial
      u(r_L')}{\partial n_L'} g_e(r_L', r'_e ; \omega) \, d r_L'.
  \end{aligned}
\end{equation}

We assume that $u = G^r$ is the solution to the homogeneous problem
$(\omega - \hat{H}(r_e))G^r(r_e, r' ; \omega) = 0$ for $r_e \in
\Omega_L$.  Since $g_e = 0$ on $\partial \Omega_L$ we have by
Eq.~(\ref{eq:represe})
\begin{equation}
  \label{eq:representation}
  \begin{aligned}
  G^r(r_e' ,r' ; \omega) =& \int_{\partial \Omega_L} \frac{1}{2} 
  G^r (r_L', r' ; \omega)
  \frac{\partial g_e (r_L' , r_e' ; \omega)}
  {\partial n_L'} \, dr_L',  \\ &\quad r_e' \in \Omega_L.
  \end{aligned}
\end{equation}
Now the representation formula (\ref{eq:representation}) can be used
to supply the boundary condition to
Eq.~(\ref{eq:variational_formulation}).  Differentiating
Eq.~(\ref{eq:representation}) with respect to $r_e'$ and letting $r_e'
\to r_L \in \partial \Omega_L$ we obtain the term corresponding to the
left boundary $\partial \Omega_L$ in
Eq.~(\ref{eq:variational_formulation}) as
\begin{equation}
  \label{eq:dton}
\begin{aligned}
 & \int_{\partial \Omega_L} \frac{1}{2} \frac{\partial G^r(r_L, r' ;
    \omega) } {\partial n_L} v(r_L) \, d r_L \\
=&  \int_{\partial \Omega_L} \int_{\partial \Omega_L} \frac{1}{2} 
  G^r(r_L', r' ; \omega) \frac{\partial^2
    g_e (r_L', r_L ; \omega)}
  {\partial n_L \partial n_L'}\, \\ 
  & \times v (r_L) \, dr_L' d r_L \\
    =&  <\hat{\Sigma}_L G^r,v>.
  \end{aligned}
\end{equation}
Here we have derived the variational form for the self-energy operator
$\hat{\Sigma}_L$. It includes line integrals over the boundary $\partial
\Omega_L$ together with a trace mapping from functions on $\Omega$ to
the functions on $\partial \Omega_{L}$. The function $\Sigma^r_L$ in
Eq.~(\ref{gammat}) is given by 
\begin{equation}
\Sigma^r_L(r_L,r_L') = \frac{1}{2} 
\frac{\partial^2 g_e (r_L', r_L ; \omega)} {\partial n_L \partial n_L'}
\end{equation}
with zero extension outside the boundary $\partial \Omega_L$.

The mapping generated above by Eq.~(\ref{eq:dton}) is called the
Dirichlet-to-Neumann mapping since in general it maps the Dirichlet
datum $u$ of a solution to a partial differential equation to the
corresponding Neumann datum $\frac{\partial u}{\partial n}$.

%-----------------------------------------------------------------

\subsection{\label{sec:fem_discretization}Finite-element
  discretization}

To obtain a numerical approximation for the Green's function $G^r$ in
the computational domain $\Omega$ we select a finite-dimensional space
$S_h$ defined on $\Omega$ and project our problem of
Eq.~(\ref{eq:variational_formulation}) into $S_h$ by solving for
$G^r_h \in S_h$ such that
\begin{equation}
  \label{eq:fem_formulation}
  \begin{aligned}
    \int_\Omega &\Big{\{} - \frac{1}{2} \nabla
    G_h^r (r,r';\omega) \cdot
    \nabla v_h(r)  \\ 
    &+ \big{[} \omega- V_{eff}(r) \big{]} G_h^r(r,r' ; \omega) v_h(r) \Big{\}} \, dr \\
    &+ <\hat{\Sigma}_L G_h^r, v_h> + <\hat{\Sigma}_R G_h^r, v_h> \\
    = &\, v_h(r')
  \end{aligned}
\end{equation}
for every $v_h \in S_h$\cite{braess}. A matrix equation is obtained by
selecting a basis $\{ \phi_i \}_{i=1}^N$ for $S_h$ and expanding
$G_h^r$ in the basis,
\begin{equation}
  G_h^r (r,r') = \sum_{i,j=1}^N g^r_{ij} \phi_i (r) \phi_j (r').
\end{equation}
Selecting $v_h = \phi_k$ in Eq.~(\ref{eq:fem_formulation}) we obtain
\begin{equation}
  \begin{aligned}
  \sum_{i,j=1}^N & g^r_{ij} \phi_j(r') \Big{\{} \int_\Omega \Big{[}
   - \frac{1}{2} \nabla \phi_i(r) \cdot
  \nabla \phi_k(r) \\ 
  &+ \big{[} \omega - V_{eff}(r) \big{]} \phi_i(r) \phi_k (r) \Big{]}
  \, dr  \\ 
  &+ < \hat{\Sigma}_L \phi_i, \phi_k> + <\hat{\Sigma}_R \phi_i, \phi_k> \Big{\}} \\
    =&  \phi_k (r').
  \end{aligned}
\end{equation}
Denoting
\begin{equation}
  \label{eq:matrix_entries}
  \begin{aligned}
  a_{ik} =& \int_\Omega \Big{(} - \frac{1}{2} \nabla \phi_i(r) \cdot
  \nabla \phi_k(r) \\ 
  & + \big{[} \omega - V_{eff}(r)\big{]} \phi_i(r) \phi_k (r) \Big{)} \, dr \\
  &+ <\hat{\Sigma}_L \phi_i, \phi_k> + <\hat{\Sigma}_R \phi_i, \phi_k>,
  \end{aligned}
\end{equation}
and exploiting the symmetry of the coefficients $g_{ij}$ we see that
$g_{ij}$'s are the entries in the inverse of the matrix the given by
Eq.~(\ref{eq:matrix_entries}).

We connect $\Sigma_{L/R}$ to the discretized forms as
\begin{equation}
\Sigma_{L/R,i,j} = <\hat{\Sigma}_{L/R} \phi_i,\phi_j >.
\end{equation}
Further, let us denote
\begin{equation}
G_h^a = \sum_{k,l} g^a_{kl} \phi_k(r) \phi_l(r'),
\end{equation}
and
\begin{equation}
\hat{\Gamma}_{L/R} = 2 \Im(\hat{\Sigma}^r_{L/R}),
\end{equation}
with
\begin{equation}
\Gamma_{L/R,ij} = <\hat{\Gamma}_{L/R} \phi_i,\phi_j> = \Gamma_{L/R,ji},
\end{equation}
since $\hat{\Gamma}_{L/R}$ is symmetric.  Now, for example, the electron
tunneling probability of Eq.~(\ref{tunneling_probability}) can be written
in the discretized form as
\begin{equation}\label{tunnelointi_dis}
\begin{aligned}
T(\omega) =& \sum_{i,j,k,l=1}^N \int_{\partial \Omega_L} \int_{\partial \Omega_L}
\int_{\partial \Omega_R}\int_{\partial \Omega_R} \\
 & \quad  \, \Gamma_L(r_L,r'_L) \, 
 g^r_{ij} \, \phi_i(r'_L) \, \phi_j(r_R) \\
&  \times \Gamma_R(r_R,r'_R) \,
g^a_{kl} \, \phi_k(r'_R) \, \phi_l(r_L) \\
& \times \, dr_L \, dr_L \, dr_R \,dr'_R, \\
=& \sum_{i,j,k,l=1}^N <\hat{\Gamma}_L \phi_i,\phi_l>  
 g^r_{ij}  < \hat{\Gamma}_R \phi_k, \phi_j>
g^a_{kl} \\
=&  \sum_{i,j,k,l=1}^N\Gamma_{L,li} \,  
 g^r_{ij} \, \Gamma_{R,jk} \, g^a_{kl}.
\end{aligned}
\end{equation}

\subsection{\label{sec:fem_basis}Finite-element basis}

\begin{figure}[htb]
\begin{center}
\epsfig{file=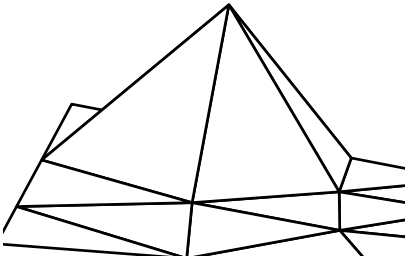,width=0.3\textwidth}
\end{center}
\caption{\label{kantafunktio} A linear basis function $\phi$. The
  function is one in a given mesh node and descends linearly to zero in
  the adjacent nodes.}
\end{figure}

So far we have not touched the subject of selecting the basis
functions $\phi_i$ in Sec. B above and thus the space $S_h$. In
principle, we could select any computable set $\{ \phi_i \}_{i=1}^N$,
but adhere to a traditional choice in the finite-element
practice, namely to the set of piecewise polynomial functions. The
basis functions are constructed as follows. Assume that $\Omega$ is
partitioned into a simple mesh of $N$ nodes and $M$ polygons $T_i$
conforming to the usual requirements imposed on a finite-element
mesh. These polygons can have a variety of shapes but the simplest
choice of triangles in two (and tetrahedral in three) dimensions will
serve our purposes. We choose the basis functions $\phi_i$ to be
element-wise linear functions that have the value one in a single node
of the mesh and zero in other nodes (see Fig.~\ref{kantafunktio}). The
corresponding finite-element space $S_h$ is
\begin{equation}
  \begin{aligned}
  S_h =& \{ v_h = \sum_{i=1}^N c_i \phi_i \, | \, c_i \in \mathbb{C} \} \\
  =& \{ v_h \in C(\Omega) \, | \, {v_h}_{|T_i} \in {\mathcal{P}}_1
  (T_i) \},
  \end{aligned}
\end{equation}
where $C(\Omega)$ denotes the set of continuous functions in $\Omega$
and ${\mathcal{P}}_1 (T_i)$ is the set of polynomials of degree one in
the polygon $T_i$.

An element-wise polynomial basis has several advantages. First,
polynomials are fast to evaluate and they can be integrated exactly on
a suitable reference element. Second, the piecewise nature of the
functions ensures that the matrix $(a_{ij})_{i,j=1}^N$ is very sparse.
Third, the accuracy of the discretization can be controlled via mesh
refinements and coarsening.

The piecewise nature of the basis functions gives rise to a
sparse matrix. Due to recent developments in linear algebra there are
fast direct solvers\cite{liu} (also parallel)\cite{mumps,gupta} for
sparse systems arising from discretization of partial differential
equations.  Since we must solve for all the coefficients $g_{ij}$ of
the approximate Green's function $g_h$ we are faced with the problem
of solving $N$ linear systems with different right-hand sides. This
kind of setting is favorable to direct methods over iterative ones.
Nevertheless, the computation itself is a time-consuming procedure and
cannot be substantially accelerated with the techniques known today.

\subsection{\label{sec:mesh_generation}Mesh generation}

An important property affecting the quality of the
finite-element approximation is the underlying mesh and especially the
shape and the size of individual elements. Several techniques for mesh
generation in two and three dimensions are available. All the
techniques have in common that they try to produce meshes with elements
of desired local size and high quality. There are also several
indicators for evaluating the quality of the shape of a single
element. Perhaps the most common is to require that there are no large
angles in the element. Typically, the larger the maximal angle of
an element is, the worse the resulting approximation will be.

%\subsubsection{\label{sec:delaunay}Delaunay meshes}

In this work we use Delaunay meshes \cite{bern_eppstein} for
triangular elements in two-dimensional problems. They are known to be
very robust in producing high-quality triangular meshes for different
shapes of domains. A Delaunay mesh can be characterized as follows. A
mesh consisting of $N$ nodes and $M$ triangular (or tetrahedral)
elements satisfies the Delaunay criterion if the circumscribe $C_j$ of
a triangle (or tetrahedron) $T_j$ of the mesh contains no nodes of the
mesh. Meshes satisfying the Delaunay criterion are called Delaunay
meshes.

It can be shown that for a given set of points in a plane a Delaunay
triangulation always exists and is even unique with a minor assumption
on the placement of the nodes. Furthermore, among all triangulations
of the nodes, the Delaunay triangulation maximizes the minimum angle
present in the triangulation. The max-min property can be usually
considered as a guarantee of high-quality elements.

%\subsubsection{\label{sec:higher_dim}Higher dimensions}

Unfortunately the Delaunay criterion is not sufficient for a high
quality tetrahedral mesh in three dimensions. This is due to the
presence of ``slivers'' in Delaunay meshes. These elements can have
very large angles deteriorating the approximation capabilities, and
yet they satisfy the Delaunay property. Therefore alternative
techniques must be sought for when producing meshes in three
dimensions. Typical approaches use a mixture of different methods,
e.g. octree methods, advancing front methods, and Delaunay methods.

However, it should be noted that the quality of the resulting mesh
produced by a mesh generation algorithm depends heavily on the
shape of the domain to be meshed. Very simple domains such as cubes
and other rectangular domains are usually well treated by virtually
any method, whereas more complicated domains having holes and cuts
need more attention.

%\hline
%\begin{multicols}{2}

%---------------------------------------------------------------
\subsection{Exterior Green's function}\label{greenulko}

The exterior Green's function for the semi-infinite leads can be
calculated numerically as the surface Green's function of a periodic
system \cite{wavelet}. In the present work the
potential is uniform in the leads along the lead axis. Therefore we
can solve for the isolated Green's function using the analytic
one-dimensional solution along the lead and the numerical transverse
wave functions $\chi_m(y)$ \cite{datta}.  The ensuing exterior Green's
function for the quasi-two-dimensional semi-infinite wire is

\begin{equation}\label{summation}
g_e = \sum_{m=1}^\infty \frac{-i \chi_m(y) \chi_m^*(y')}{k_m} \left(
e^{ik_m(x-x')} - e^{ik_m(x+x')} \right),
\end{equation}
where $\chi_m(y)$'s are 
solutions to the Kohn-Sham equation
\begin{equation}\label{poikki}
\left( -\frac{1}{2}\nabla^2 -V_{eff}(y) \right) \chi_m(y) = \epsilon_m
\chi_m(y),
\end{equation}
with
\begin{equation}
k_m = \sqrt{2(\omega - \epsilon_m).}
\end{equation}
We solve Eq.~(\ref{poikki}) using self-consistency iterations for the
electron density and the potential profile $V_{eff}(y)$.  As explained
before we use a model in which the positive charge forms a thin wire
and the electron wave functions spread out of this charge.  The
effective potential $V_{eff}$ consists only of $V_{xc}$ and $V_c$, and
no external potential is applied. In practice the summation in
Eq.~(\ref{summation}) is  truncated typically after a few tens of
states so that the results are well-converged.

\begin{figure}[htb]
\begin{center}
\epsfig{file=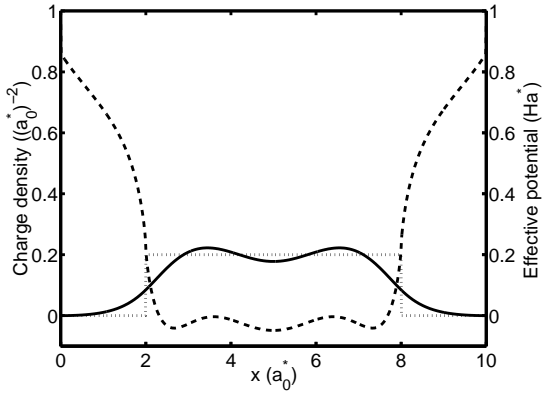,width=0.4\textwidth}
\end{center}
\caption{\label{reunaehto} Electron density (solid line), positive
  background charge (dotted line) and $V_{eff}$ (dashed line) for an
  infinite uniform wire.}
\end{figure}

The charge densities resulting from this calculation are used in the
boundary conditions when calculating the Coulomb potential of the
nanosystem. The total charge per unit length is zero in an infinite
wire, but there are local variations in the charge density in the
transverse direction. As an example, we show in Fig.~\ref{reunaehto}
the effective potential and the positive and negative charge densities
in a case with two transversal modes in the wire. A cut perpendicular
to the wire axis is shown.

%---------------------------------------------------------------
\subsection{Coulomb interactions}

The effective potential is also calculated using the FEM and the same
mesh as for the Green's functions is used. $V_{xc}$ is simply
evaluated in every node point. The potential charge densities are
two-dimensional but the Coulomb is treated in three dimensions. In
this case it is not efficient to solve for the three-dimensional
Poisson equation, but to evaluate the integral
\begin{equation}
V_c(r) = \int \frac{ \rho(r')- \rho_p(r')}{ |r-r'|}dr'.
\end{equation}
Above, $\rho$ is the electron density and $\rho_p$ is the positive
background charge density. The integral is evaluated by integrating
basic functions in every element. For elements with no pole ($r$ is
not inside the element), the integral is evaluated using the Gaussian
quadrature rules for triangles \cite{kolmioGaus}. Elements which have
$r$ in one corner are evaluated by making a mapping from the triangle
to a square in which the pole disappears \cite{nelio}.

%---------------------------------------------------------------
\section{Test systems}\label{test}

This section is devoted for testing and demonstrating our scheme.
First the transmission probability over a given potential well and
through a given bottle-neck potential are determined. The aim of these
non-self-consistent calculations is to provide, trough the comparison
with the exact results, an idea of the numerical accuracy of our
methods.  Thereafter we demonstrate the possibilities of the scheme by
solving self-consistently the electronic structure and the current
under a bias voltage for different resonant tunneling systems.

\subsection{Transmission probability over a potential well}

\begin{figure}[htb]
\begin{center}
\epsfig{file=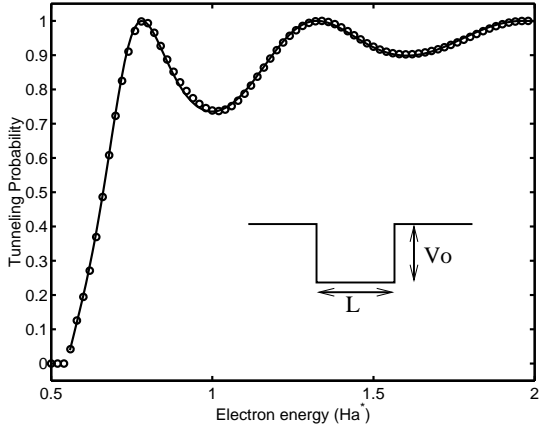,width=0.4\textwidth}
\end{center}
\caption{\label{kuoppa} Transmission probability over a potential
well. The solid line corresponds to the analytic solution of
Eq.~(\ref{kuoppayhtalo}) and the circles are calculated using the FEM
code. In this calculation $L=10 \, a_0^*$, $V_0 = 1 \, {\rm Ha}^*$,
the width of the wire $W=3 \, a_0^*$, and the average distance between
the FEM mesh nodes $h=0.3 \, a_0^*$.}
\end{figure}

Basic quantum mechanics gives the transmission probability over a
potential well (see the inset in Fig.~\ref{kuoppa}) as
\begin{equation}\label{kuoppayhtalo}
T(\omega) = 2\left[ 1 + \frac{V_0^2 \sin^2(\sqrt{2(\omega + V_0)} L)}{4
\omega ( \omega + V_0) }\right]^{-1},
\end{equation}
where $V_0$ and $L$ are the depth and the length of the well,
respectively, and $\omega$ is the electron energy. Our numerical approach
obeys this result accurately. For example, Fig.~\ref{kuoppa} gives the
transmission probability calculated using
Eqs.~(\ref{tunneling_probability}) and (\ref{tunnelointi_dis}) for a
narrow wire with a potential well. For the energies shown there is
only one transverse mode in the wire. The good agreement between the
numerical and analytic results indicates that the FEM mesh is fine
enough.

%---------------------------------------------------------------

\subsection{Transmission probability through a bottle-neck potential}

\begin{figure}[htb]
\begin{center}
\epsfig{file=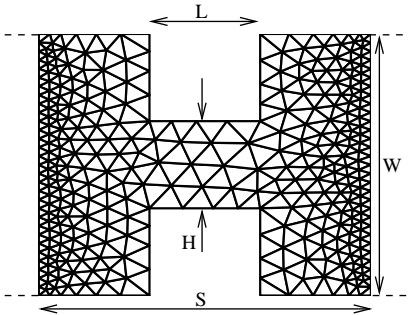,width=0.3\textwidth}
\end{center}
\caption{\label{kavennus} Bottle-neck model potential. The potential
is constant inside the leads and in the bottleneck between the
leads. At the boundaries the potential rises to infinity. The
dimensions are $L =H =10\,a_0^*$ and $W = 30 \, a_0^*$. The length of
the calculation area $S= 30\, a_0^*$. The FEM mesh shown has smaller
elements near the boundaries $\partial \Omega_{L/R}$.}
\end{figure}

\begin{figure}
  \centering \mbox{\subfigure
  {\epsfig{figure=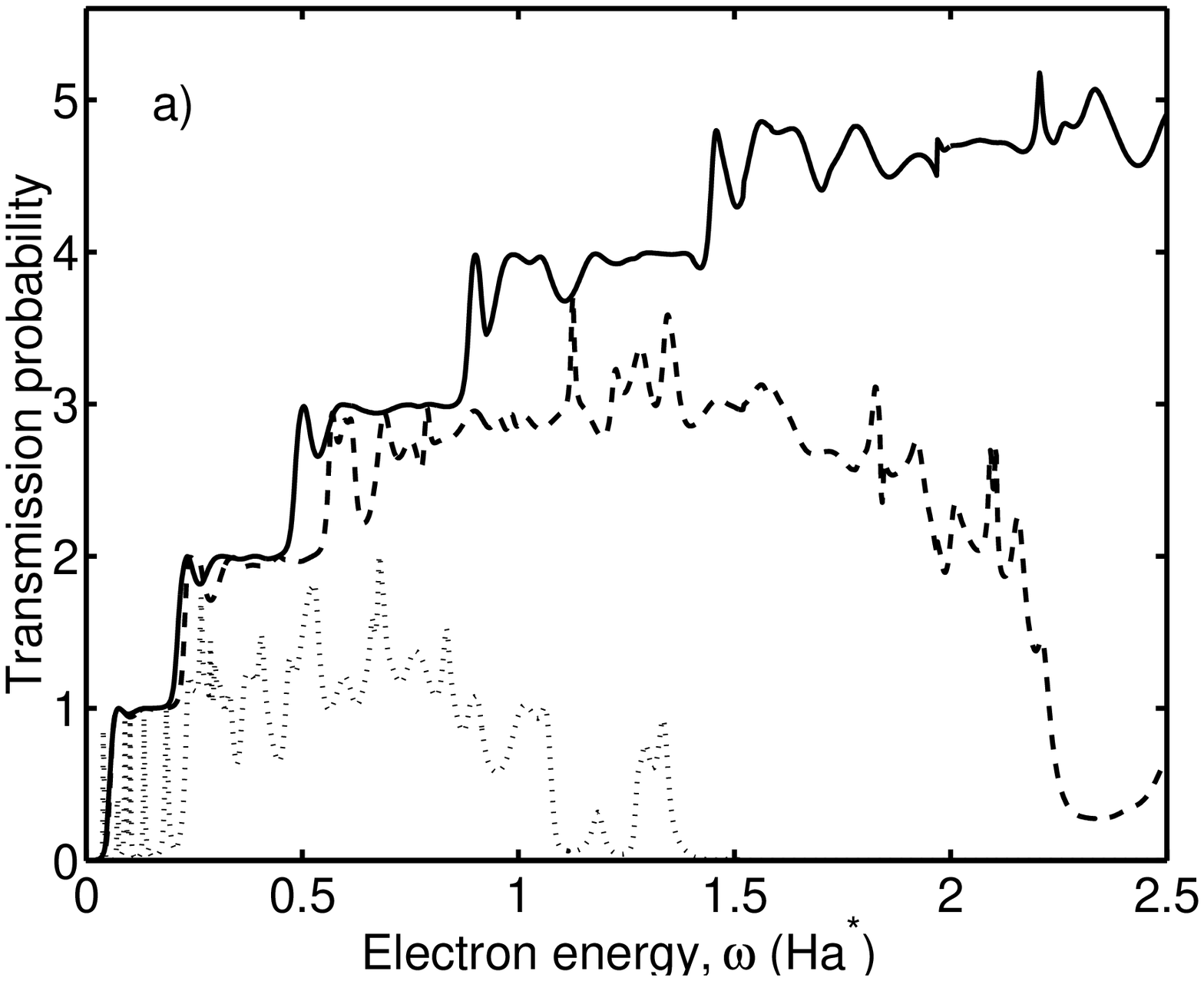, width=0.4\textwidth}} } \\ \mbox{
  \subfigure
  {\epsfig{figure=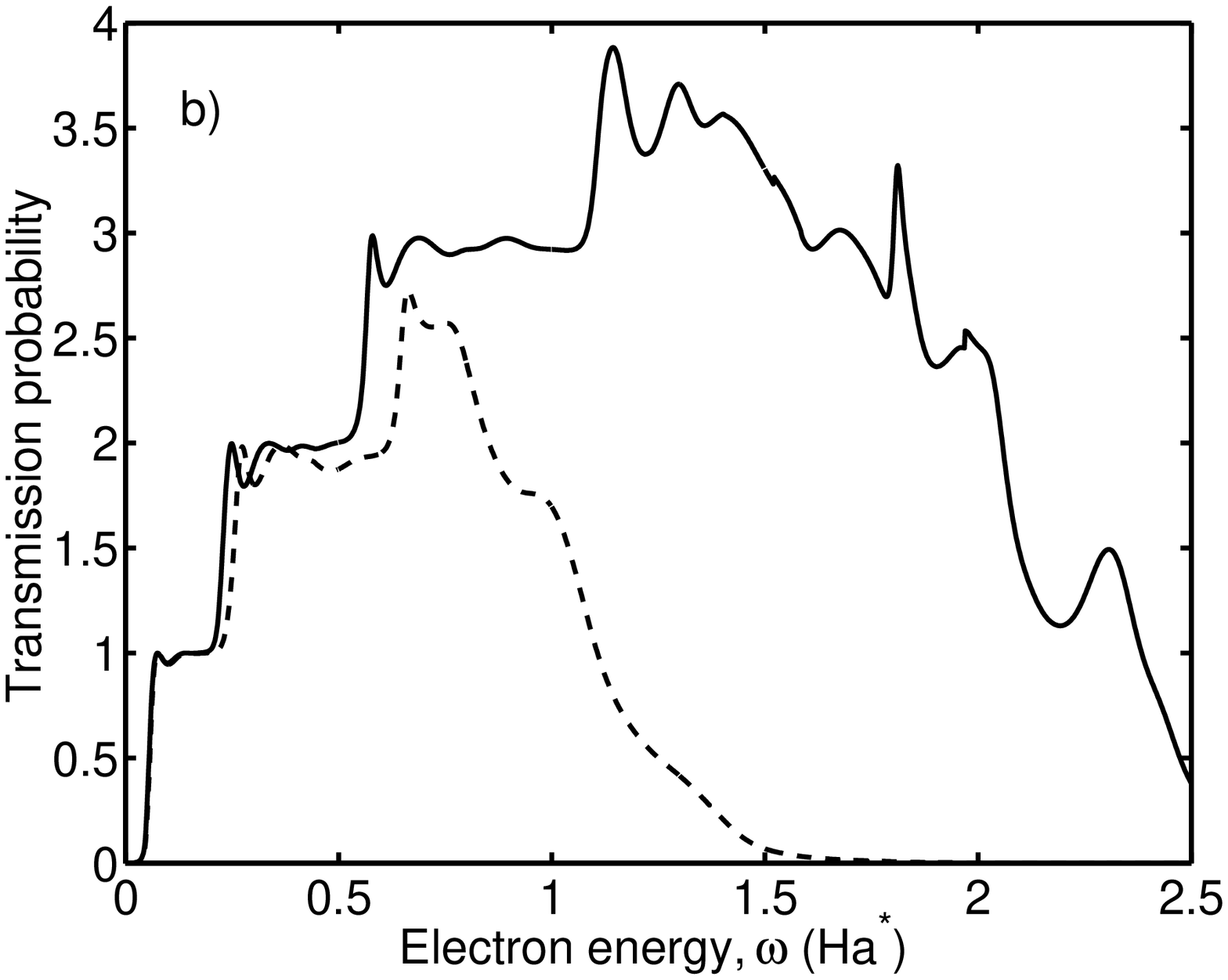, width=0.4\textwidth}} }
\caption{\label{tarkkuus} Electron transmission probability as a
function of the energy for different FEM meshes. a) All the elements
in each calculation are of the same size. The FEM node distance $h=1\,
a_0^*$ (solid line), $h=2 \, a_0^*$ (dashed line) and $h=3 \, a_0^*$
(dotted line). b) The elements are smaller near the boundaries
$\partial \Omega_{L/R}$ (see Fig.~\ref{kavennus}). The minimum
distance $h_{min} = 1 \, a_0^*$ and the maximum distance $h_{max} = 2
\, a_0^*$ (solid line) and $h_{max} = 3 \, a_0^*$ (dashed line).  }
\end{figure}

Next we study how the FEM node density affects the results. We
calculate the electron transmission probability as a function of
energy using different FEM meshes. Our scattering potential is a
bottleneck shown in Fig.~\ref{kavennus}.  The electron transmission
probability is shown in Fig.\ref{tarkkuus} as a function of the
energy. In stepwise jumps in the transmission probability mean that
new transverse modes emerg with increasing energy $\omega$. The narrow
peaks near the beginning of each step correspond to the constructive
interference of the incident wave with the wave reflected twice at the
lead-bottleneck boundaries \cite{kavennus}. Increasing the energy
means making the electron wavelength shorter so that more points are
needed to describe the wave functions. Thus, with a fixed element size
$h$ it is possible to characterize transversal modes up to a certain
energy only. Thereafter the transmission probability collapses due
to the a loss of numerical stability.

In Fig. \ref{tarkkuus}a the size of the elements in each calculation
is the same throughout the whole calculation area.  According to the
two uppermost curves corresponding to the FEM node distances  $h= 1 \,
a_0^*$ and $h=2\, a_0^*$, we need about 4 nodes between the adjacent
zero-value lines of the electron wave function. This means that the
FEM node distance of $h=3 \, a_0^*$ should give a reasonable result
for the first transversal mode. In contrast, the results show large
oscillations of the transmission due to discretization errors.  The
reason for this is that the pole of the real part of the Green's
function is not approximated accurately enough.  When determining the
transmission the arguments of the Green's function are on the opposite
boundaries (Eq.~(\ref{tunneling_probability})). These Green's function
values are calculated by solving a linear equation problem in which
one of the arguments of $G^r(r,r')$ is fixed e.g. on the left
boundary, $\partial \Omega_L$ and the other argument runs over the
central region to the right boundary $\partial \Omega_R$. If the FEM
mesh is not dense enough near the left boundary where the pole is a
large numerical error propagates to the elements needed in
Eq.~(\ref{tunneling_probability}) \cite{napa_virhe}.  In
Fig.~\ref{tarkkuus}b the number of points at the boundaries $\partial
\Omega_{L/R}$ is larger than inside the calculation area $\Omega$. The
figure shows that the effects of the discretization errors are now
strongly reduced at low energies, but the transmission probability at
high energies collapses as fast as in Fig.~\ref{tarkkuus}.  In
conclusion, when one wants to describe the transmission probability
only up to a certain energy value, the optimum way to choose the sizes
of the elements is to use smaller elements near the boundaries
$\partial \Omega_{L/R}$ than inside the area $\Omega$. In this simple
test system the bottleneck potential is relative wide, but if the
bottleneck is narrow in comparison with to the rest of the wire, it is
reasonable to refine the mesh also in the neck region. Finally, the
above refinement is also needed when calculating the electron density
in nonequlibrium using Eq.~(\ref{ele2}). The real part of $G^r(r,r')$
is needed between a point on the boundary, $\partial \Omega_{L,R}$ and
an arbitrary point in the central region $\Omega$.

%---------------------------------------------------------------

\subsection{Resonant tunneling through double-barrier potential systems}

\subsubsection{Symmetric barrier system}

\begin{figure}[htb]
\begin{center}
\epsfig{file=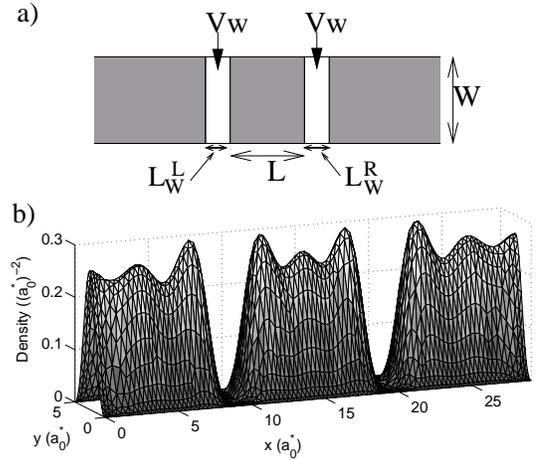,width=0.4\textwidth}
\end{center}
\caption{\label{kaksoisvalli} Double-barrier potential system.  a) The
model. The gray areas correspond to the positive background charge. At
the gaps there is an additional potential $V_w=2 \, \rm{Ha}^*$. The
size of calculation area $\Omega$ is $29 \times 5 \, (a_0^*)^2$, the
width of the background charge $W= 3 \,a_0^*$ and length of the
quantum dot $L=9 \, a_0^*$. Case A has $L_W^{L/R} = 1 \, a_0^*$ and
case B $L_W^{L/R} = 1.25 \, a_0^*$. The number of FEM nodes used in
the calculations is 2105. b) The total electron density at zero
bias voltage for case A.}
\end{figure}

In this subsection we demonstrate the potential of our scheme by
showing results of self-consistent electronic-structure calculations
for 2D nanostructures under a finite bias voltage. We restrict
ourselves to zero temperature calculations. The test system is a
double-barrier potential structure, a schematic sketch of which is
shown Fig.~\ref{kaksoisvalli}a. A jellium wire is cut by two vacuum
regions and additional potential barriers are introduced within them
in order to adjust the potential and the transmission. We consider two
special cases. Case A has thinner potential walls $L_W^{R/L} = 1 \,
a_0^*$ than case B for which $L_W^{R/L} = 1.25 \, a_0^*$. This
difference means that the connection to the leads differs remarkably
its the strength.  We make contact with real semiconductor systems
by converting our results from the effective atomic units to the
SI-units using the effective mass of electrons $m^* = 0.067$ and the
dielectric constant $\epsilon = 12.4$ for GaAs.  Then $a_0^* = 9.779 \,
\rm{nm}$ and $1 \, \rm{Ha}^* = 11.8672 \, \rm{meV}$.  The
positive background charge density $0.2 \, (a_0^*)^{-2} \approx 2
\cdot 10^{15} \, \rm{m}^{-2}$ corresponds to a reasonable electron
density at the GaAs/AlGaAs interface.  The groundstate electron
density of the double-barrier system is shown in
Fig.~\ref{kaksoisvalli}b, exhibiting Friedel
oscillations in both leads. The wires are so thin that only one
transverse mode is occupied.

\begin{figure}[htb]
  \centering \mbox{\subfigure
  {\epsfig{figure=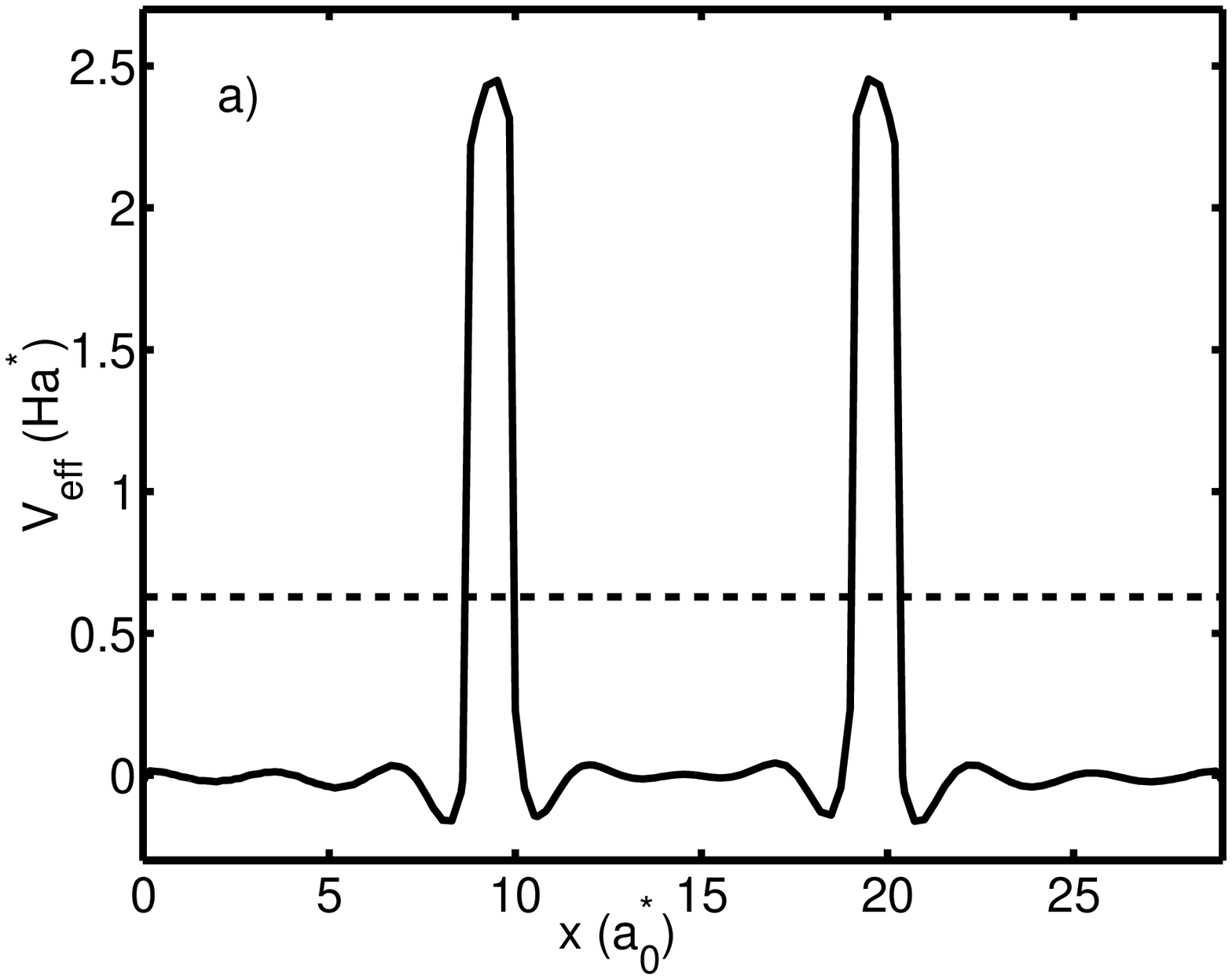, width=0.4\textwidth}} } \\ 
  \mbox{
  \subfigure
  {\epsfig{figure=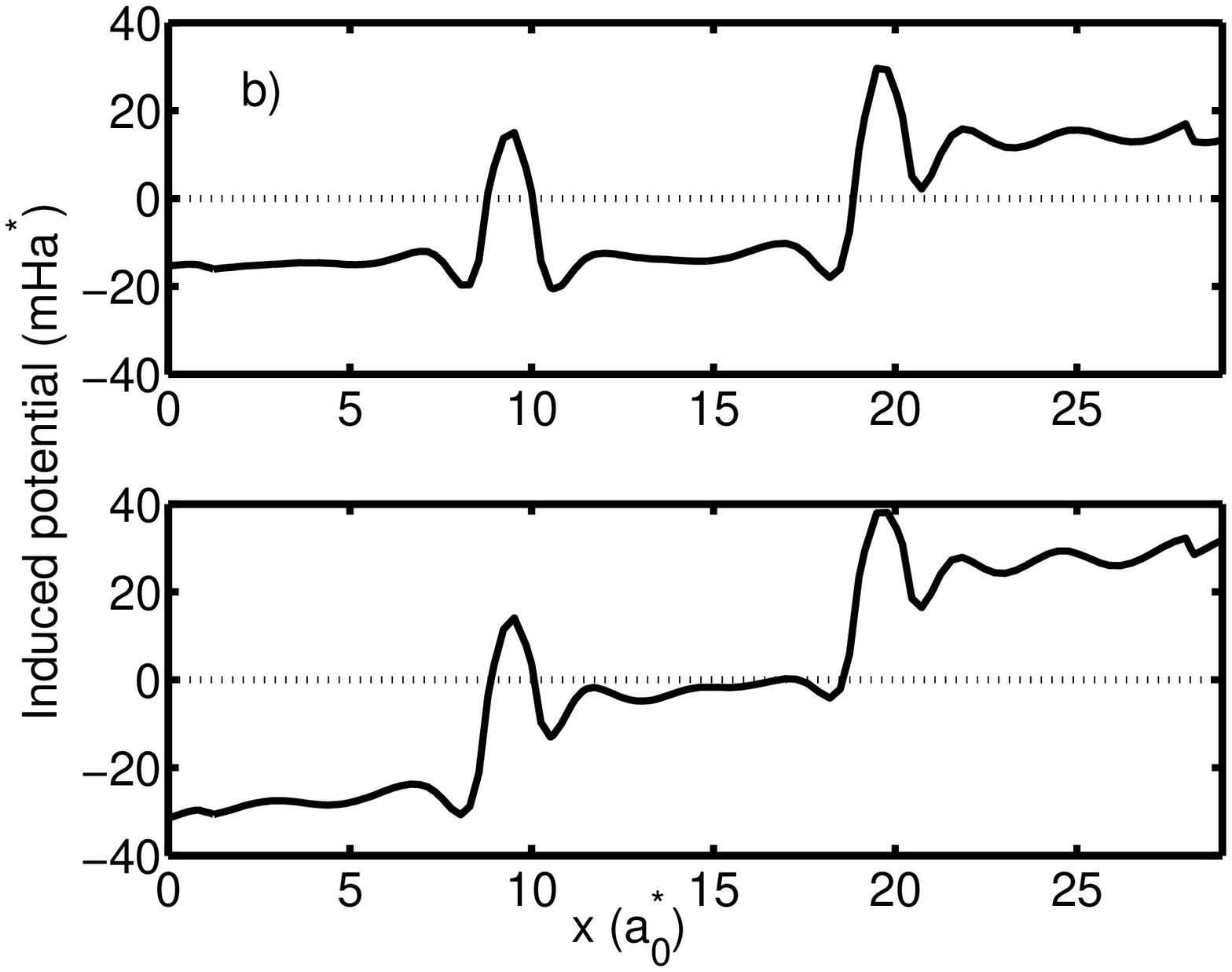, width=0.4\textwidth}}}
 \caption{\label{heff_poikki} Double-barrier potential system B. a)
  The zero-bias voltage effective potential along the symmetry
  axis. The energy-zero corresponds to the bottom of energy band in an
  infinite 2D-system with the electron density of
  $0.2\,(a_0^*)^2$. The Fermi level is shown by the dashed line.  b)
  The change of $V_{eff}$ due to bias voltage. In the upper panel
  $\Delta V_{bias} = 0.03 \, Ha^*$ (0.36 meV) and lower panel $\Delta
  V_{bias} = 0.06 \, Ha^*$ (0.71 meV).}
\end{figure}

The effective potential along the symmetry axis of the double-barrier
system at zero bias voltage is shown in Fig.~\ref{heff_poikki}a. The
potential barriers are so small that the quantum dot is strongly
connected to the leads. When we add the bias voltage to the system,
the potential of right lead increases and that of the left lead
decreases. The change of $V_{eff}$ for case B is shown in
Fig.~\ref{heff_poikki}b.  The maximum bias voltage applied is small in
comparison to the barrier heights. The potential drop occurs between
the potential walls, not in the leads. This is expected because the
leads are ballistic, with no scatterers at all. At small $\Delta
V_{bias}$ values the potential in the quantum dot stays at the level
of the potential in the left lead. This is seen in the upper panel of
Fig.~\ref{heff_poikki}b. When $\Delta V_{bias}$ is large enough the
potential in the dot rises close to the mean value in the leads (see
the lower panel).  A nearly inversion-symmetric potential develops. In
case A the potential in the quantum dot develops differently. It
follows mainly the potential level of the right lead for all bias
voltages studied.

\begin{figure}
  \centering \mbox{\subfigure
  {\epsfig{figure=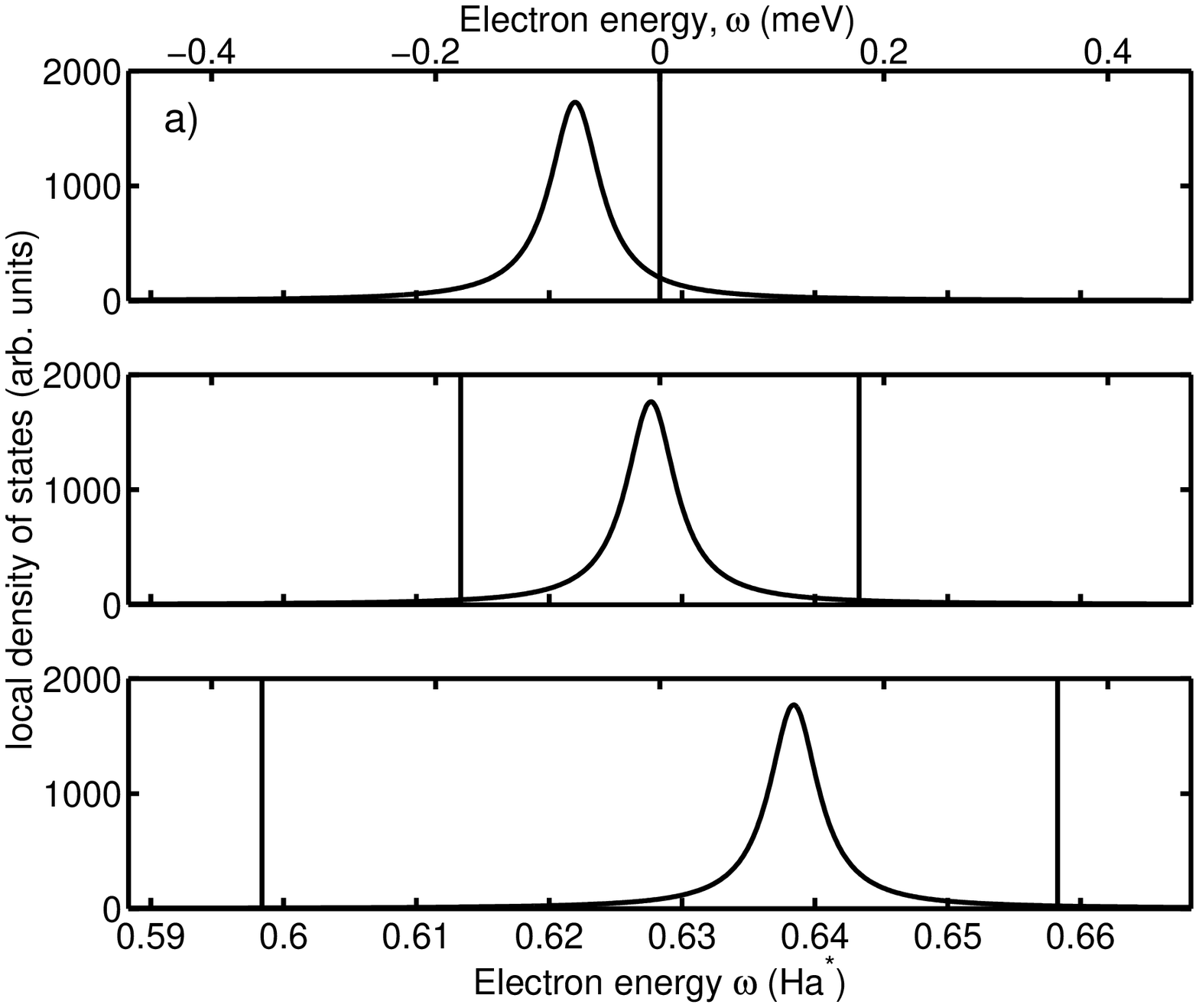, width=0.4\textwidth}} } \\ 
  \mbox{
  \subfigure
  {\epsfig{figure=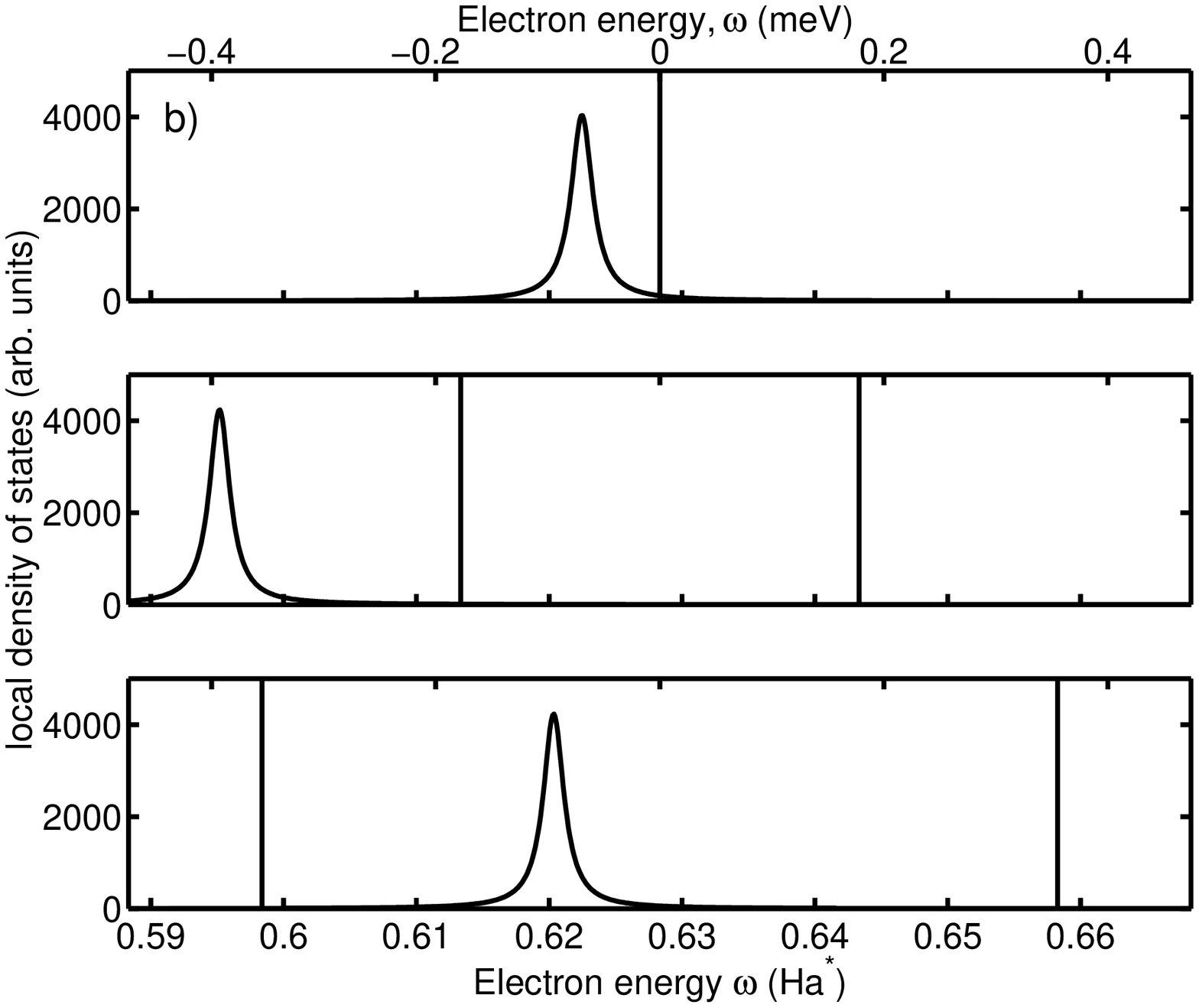, width=0.4\textwidth}} }
\caption{\label{tilatiheys} LDOS in the region between the barriers
shown in Fig.~\ref{kaksoisvalli}. a) LDOS for case A with
narrow barriers. b) LDOS for the case B with wide barriers.  The
vertical lines denote the Fermi level position in the leads. Both in a)
and b) the uppermost  panels correspond to the zero-bias calculation, the
middle panels to $\Delta V_{bias} = 0.03 \, Ha^*$ whereas the
lowest panels correspond to $\Delta V_{bias} = 0.06 \, Ha^*$.}
\end{figure}

The behavior of the potential level in the quantum dot is connected to
the occupation of the dot resonance state and its position relative to
the lead Fermi levels.  Fig.~\ref{tilatiheys} shows the local density
of states (LDOS) calculated by integrating over the quantum dot area.
For the zero bias voltage, both cases, A and B, have a resonance peak
below the Fermi level.  When the bias $\Delta V_{bias}$ is applied the
potentials and the Fermi levels are shifted by $+ \frac{1}{2}\Delta
V_{bias}$ and $- \frac{1}{2}\Delta V_{bias}$ in the left and right
leads, respectively. This defines the so-called bias window on the
energy axis. At small $\Delta V_{bias}$ the value the resonance peak
to case B moves down in energy. The resonance, which gives a large
contribution to the charge in the dot is below the left Fermi
level. The bias induced charge redistribution takes place near the
left barrier. Thus the potential in quantum dot stays at the level of
the left lead. However, when $\Delta V_{bias}$ is large enough the
resonance peak enters the bias window, the charge redistribution
occurs quite symmetrically at both barriers and the potential level in
the quantum dot is in the middle between the left and right lead
levels. The resonance peak of case A is wider than that of case B
because the connection to the leads is stronger.  The wide resonance
enters the bias window at a low bias value and its position follows
the Fermi level of the right lead. Then the bias-induced charge
redistribution takes place at the left barrier and the potential level
in the dot follows that in the right lead. The asymmetric behavior of
the voltage drop in our model systems has analogies with the case of
atomic chains between two electrodes \cite{siesta3}

\begin{figure}[htb]
\begin{center}
\epsfig{file=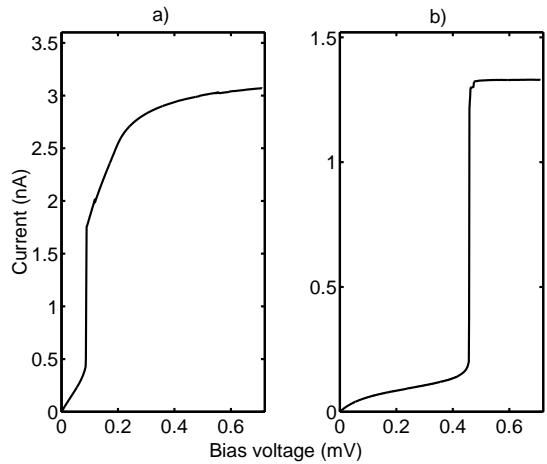,width=0.4\textwidth}
\end{center}
\caption{\label{virta_jannite} Current as a function of the bias
 voltage for the double-barrier potential systems shown
 Fig.~\ref{kaksoisvalli}. a) Case A with the barrier width of $1 \,
 a_0^*$. b) Case B with the barrier width of $1.25 \, a_0^*$. The
 zero-bias conductivities of case A and B are $0.060 \, G_0$ and
 $0.014 \, G_0$, respectively. }
\end{figure}

The position of the resonance peak relative to the Fermi levels has a
large effect on the electron transmission probability through the
double-barrier potential system. The current flow is due to the states
with energies between right and left Fermi-levels i.e. in the bias
window. When the resonance peak moves into this region there is a
steep increase in the current. Thereafter the current stays
approximately constant as a function of the bias voltage. This
characteristic behavior of the double barrier potential is visible in
Fig.~\ref{virta_jannite}.  Case B with the sharper resonance peak
has a steeper raise of the current than case A. Moreover, the raise
occurs at a higher bias voltage in case B than in case A.

\subsubsection{Asymmetric barriers}

So far both the potential barriers in the system of
Fig.~\ref{kaksoisvalli}a have been identical. Inspired by the prospect
to use non-symmetric molecules as rectifiers \cite{mujica,tasa_taylor}
we have studied also double-barrier systems with non-identical
barriers. The zero-bias conductivities of the cases A and B (see
Fig.~\ref{virta_jannite} and its caption) are $0.060 \, G_0$ and
$0.014 \, G_0$. These are of the same order in magnitude as conductivities
calculated for molecules between electrodes \cite{tasa_taylor}.  In
the next example we have reduced the height of the second barrier in
case A by a factor of two in order to create an asymmetric system.

\begin{figure}[htb]
\begin{center}
\epsfig{file=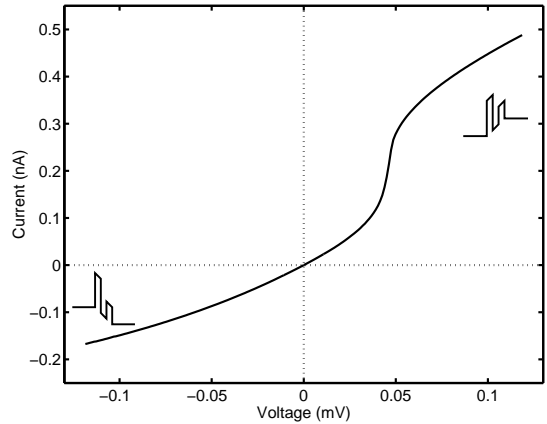,width=0.4\textwidth}
\end{center}
\caption{\label{tasa_IV} Current-voltage curve for a 
  double-barrier potential system with asymmetric barriers.}
\end{figure}

\begin{figure}[htb]
\begin{center}
\epsfig{file=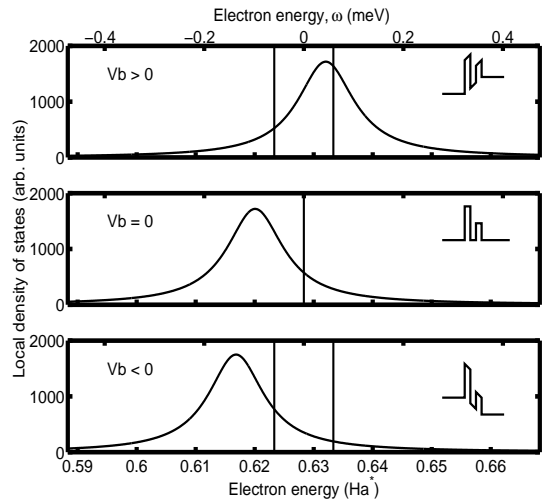,width=0.4\textwidth}
\end{center}
\caption{\label{tasa_tila} LDOS for the double-barrier potential
  system with asymmetric potential barriers. The LDOS corresponds to
  the quantum dot region between the barriers.}
\end{figure}

The ensuing current-voltage curve is shown in
Fig.~\ref{tasa_IV}. The curve is asymmetric with respect to the
direction of the applied bias.  The double-barrier system shows a
clear rectification effect resembling that for asymmetric molecular
wires \cite{tasa_taylor}.  The reason for the rectification effect is
seen in the LDOS in the quantum dot given in Fig.~\ref{tasa_tila}. When
the bias over the system is zero a resonance peak is below the Fermi
level as it was in the previous cases A and B. For positive bias
voltages (the potential is higher in the lower-barrier side) the
resonance peak moves up in energy and the resonance is emptying of
electrons. This causes the increase in the conductivity. In the case of
negative bias voltages (the potential is higher in the 
higher-barrier side) the resonance peak follows the Fermi energy of the
lower-potential lead. The situation is similar to that of system B
above at low bias. The resonance does not enter the bias window as fast as
in the case of the positive voltage and the current increases slowly.

%---------------------------------------------------------------
\section{Conclusions}

We have developed a computational scheme to model two-dimensional
nanostructures connected to two semi-infinite leads.  The electron
density and the current are calculated self-consistently using the
non-equilibrium Green's function approch. The single-particle electron
states are handled within the density-functional theory.

We have formulated the problem using the finite-element
approximation. In this approximation the boundary conditions are easy
to derive and implement. We have shown the derivation of the
Dirichlet-to-Neumann boundary conditions and the discretized forms of
physical quantities such as the tunneling probability.

Tests with model potential systems show the numerical accuracy
and its dependence on the finite-element mesh chosen. Especially, we
show that for efficient accurate calculation is important to refine
the mesh near the boundaries between central region and the
boundaries. Self-consistent calculations for resonant tunneling
structures demonstrate the efficiency of the scheme.

We have treated systems with upto 10 000 degrees of
freedom. Three-dimensional atomistic systems described by of
pseudopotentials would need roughly one order of magnitude more
degrees of freedom withch is wihin present-day computational
capabilities. The present two-dimensional work is an important step
in the development towards three-dimensional atomistic modeling of
non-equilibrium transport in nanoscale devices.

\acknowledgments

We thank Dr. Per Hyldgaard and Dr. Harri Hakula for many enlightening
discussions.

We acknowledge the generous computer resources from the Center for
Scientific Computing, Espoo, Finland.  This research has been
supported by the Academy of Finland through its Centers of Excellence
Program (2000-2005).

%------------------------------------------------------------------

\end{multicols}


\begin{references}

\bibitem{datta} S. Datta, {\it Electronic Transport in Mesoscopic
  Systems}, Cambridge University Press (1995).

\bibitem{montecarlo} W. M. C. Foulkes, L. Mitas and R. J. Needs and
G. Rajagopal, Rev. Mod. Phys. {\bf 73}, 33 (2001).

\bibitem{manninen}S. M. Reimann and M. Manninen, 
Rev. Mod. Phys. {\bf 74}, 1283 (2002). 

\bibitem{lang} N. D. Lang, Phys. Rev. B {\bf 52} 5335 (1995).

%Greenin fuktio teoriaa
\bibitem{green1} Y. Xue, S. Datta and M. A. Ratner,
  Chemical Physics {\bf 281} 151-170 (2002).

\bibitem{beck} For a review, see T. Beck, Rev. Mod. Phys, {\bf 72}, 1041
  (2000).

\bibitem{hsl} The Harwell Subroutine Library, see
http://www.cse.clrc.ac.uk/nag/hsl/

\bibitem{arias} For a review, see T. A. Arias, Rev. Mod. Phys. {\bf
  \bf 71}, 267 (1999).

\bibitem{uasp} See, for example G. Kresse and J. Furthm\"uller,
  Phys. Rev. B {\bf 54}, 11169 (1996).

% transiesta papereita
\bibitem{siesta1} J. Taylor, H. Guo and J. Wang, Phys. Rev. B
{\bf 63} 245407 (2001).

% transiesta papereita
\bibitem{siesta2} M. Brandbyge, J. Mozos, P. Ordejo\'n,
  J. L. Taylor and K. Stokbro, Phys. Rev. B {\bf 65}, 165401
  (2002).

\bibitem{bernholc} M. B. Nardelli, J.-L. Fattebert, and
J. Bernholc Phys. Rev. B {\bf 64}, 245423 (2001).

\bibitem{damle} P. S. Damle, A. W. Ghosh, and S. Datta,
Phys. Rev. B {\bf 64}, 201403(R) (2001).

%wawelet paperi
\bibitem{wavelet} K. S. Thygesen, M. V. Bollinger, and K. W. Jacobsen,
Phys. Rev. B {\bf 67}, 115404 (2003).

% FEM elektronirakenteessa
\bibitem{fem1} J. E. Pask, B. M. Klein, P. A. Sterne and C. Y. Fong,
  Computer Physics Communications {\bf 135} 1-34 (2001).

% FEM elektronirakenteessa
\bibitem{fem2} J. E. Pask, B. M. Klein, C. Y. Fong and P. A. Sterne
 Phys. Rev. B  {\bf 59}, 12352 (1999).

% FEM elektronirakenteessa
\bibitem{fem3} E. Tsuchida and M. Tsukada, Phys. Rev. B {\bf 54},
7602-7605 (1996).

% FEM elektronirakenteessa
\bibitem{fem4} E. Tsuchida and M. Tsukada Phys. Rev. B {\bf 52},
5573-5578 (1995).

% FEM elektronirakenteessa
\bibitem{fem5} S. R. White, J. W. Wilkins and M. P. Teter,
Phys. Rev. B {\bf 39}, 5819-5833 (1989).

\bibitem{femKirja} S. C. Brenner, L. R. Scott, {\it The
  Mathematical Theory of Finite Element Methods}, Second Edition,
  Springer (2002). 

\bibitem{hirose} K. Hirose, F. Zhou, and N. S. Wingreen 
Phys. Rev. B {\bf 63}, 075301 (2001).

% vaihto korrelaatio paperi
\bibitem{xc1} C. Attaccalite, S. Moroni, P. Gori-Giorgi, and G. B. Bachelet,
Phys. Rev. Lett. {\bf 88}, 256601 (2002).

% vaihto korrelaatio paperi
\bibitem{xc2}
P. Gori-Giorgi, C. Attaccalite, S. Moroni, and G. B. Bachelet,
Int. J. Quantum Chem. {\bf 91}, 126 (2003).

\bibitem{liu}J.W.H. Liu, SIAM Review, {\bf 34} 82-109 (1992)

\bibitem{mumps}P.R. Amestoy, I.S. Duff, and J.-Y. L'Excellent, Comput.
  Methods in Appl. Mech. Eng., {\bf 184}, 501-520 (2000)

\bibitem{gupta}A. Gupta, ACM Transactions on Mathematical Software,
  {\bf 28} (2002)

\bibitem{bern_eppstein}M. Bern and D. Eppstein, \textit{Computing in
    Euclidean Geometry}, 23-90, (World Sci. Publishing, River Edge,
  NJ, 1992)

\bibitem{braess}D. Braess, \textit{Finite Elements} (Cambridge
  University Press, 2001)

\bibitem{kolmioGaus} R. Cools and P. Rabinowitz, J.
  Comp. Appl. Math., {\bf 48}, 309 (1993).


\bibitem{nelio} M. G. Duffy, SIAM Journal on Numerical Analysis,
{\bf 19}, 1260-1262 (1982).


\bibitem{kavennus} A. Szafer and A. D. Stone,
  Phys. Rev. Lett. {\bf 62}, 300-303 (1989). 


\bibitem{napa_virhe}I. Babuska, T. Stroubolis, {\it The Finite
  Element Method and its Reliability}, Oxford University Press (2001).


\bibitem{siesta3} M. Brandbyge, N. Kobayashi, M. Tsukada 
Phys. Rev. B {\bf 60}, 17064-17070 (1999).

\bibitem{mujica} V. Mujica, M. A. Ratner and A. Nitzan,
  Cham. Phys. {\bf 281}, 147-150 (2002).

\bibitem{tasa_taylor} J. Taylor, M. Bradbuge, and K. Stokbro,
  Phys. Rev. Lett. {\bf 89}, 138301 (2002).



\end{references}
\end{document}